\documentclass[amsmath,amssymb,showpacs,draft,preprint,floatfix]{revtex4}
\usepackage{graphicx}
\usepackage{latexsym}
\begin{document}
\title{\bf Relation between the field quadratures and the characteristic function of a mirror}
\author{Blas M. Rodr\'{\i}guez  and H\'ector Moya-Cessa}
\affiliation{Instituto Nacional de Astrof\'{\i}sica, Optica y
Electr\'onica, Apdo. Postal 51 y 216, 72000 Puebla, Pue., Mexico }
\date{\today}
\begin{abstract}
We analyze the possibility of measuring the state of a movable
mirror by using its interaction  with a quantum field. We show
that measuring the field quadratures allows to reconstruct the
characteristic function corresponding to the mirror state.
\end{abstract}
 \maketitle

The reconstruction of a quantum state is a central topic in
quantum optics and related fields \cite{ris,leo}. During the past
years, several techniques have been developed, for instance the
direct sampling of the density matrix of a signal mode in
multiport optical homodyne tomography \cite{zuk}, tomographic
reconstruction by unbalanced homodyning \cite{wal}, reconstruction
via photocounting \cite{ban}, cascaded homodyning \cite{kis} to
cite some. There have also been proposals to measure
electromagnetic fields inside cavities \cite{lut,moy} and
vibrational states in ion traps \cite{lut,bar}. In fact the full
reconstruction of nonclassical states of the electromagnetic field
\cite{smi} and of (motional) states of an ion \cite{lei} have been
experimentally accomplished. The quantum state reconstruction in
cavities is usually achieved through a finite set of selective
measurements of atomic states \cite{lut} that make it possible to
construct quasiprobability distribution functions such as the
Wigner function, that constitute an alternative representation of
a quantum state of the field.

Recently there has been interest in the production of
superposition states of macroscopic systems such as a moving
mirror \cite{bouw}. It is therefore of interest to have schemes to
measure the non-classical states that may be generated for the
moving mirror. Here we will propose a method to relate the
quadratures of the field to the characteristic function associated
to the density matrix of the mirror.

The interaction between a quantum electromagnetic field and a
movable mirror (treated quantum mechanically)  has a relevant
Hamiltonian given by \cite{manci}

\begin{equation}
H=\hbar(\omega a^{\dagger}a + \Omega b^{\dagger}b -g
a^{\dagger}a(b^{\dagger}+b)) \label{ham}
\end{equation}
where $a$ and $a^{\dagger}$ are the annihilation and creation
operators for the cavity field, respectively. The field  frequency
is $\omega$. $b$ and $b^{\dagger}$ are the annihilation and
creation operators for the mirror oscillating at a frequency
$\Omega$ and
\begin{equation}
g = \frac{\omega}{L}\sqrt{\frac{\hbar}{2m\Omega}},
\end{equation}
with $L$ and $m$ the lenght of the cavity and the mass of the
movable mirror.

We can re-write the Hamiltonian (\ref{ham})in the form \cite{bose}
\begin{equation}
H= D_m(\eta a^{\dagger}a)\left(\omega a^{\dagger}a + \Omega
b^{\dagger}b -\epsilon (a^{\dagger}a)^2\right)D_m^{\dagger}(\eta
a^{\dagger}a)
\end{equation}
where $ \epsilon=g\eta$ with $\eta=g/\Omega$ and  the displacement
operator is given by
\begin{equation}
D_m(\beta)=e^{\beta b^{\dagger}-\beta^* b},
\end{equation}
with $N=a^{\dagger}a$. Then the unitary evolution operator is
simply
\begin{equation}
U(t)= e^{\frac{-iHt}{\hbar}}D_m(\eta N)e^{-it\left(\omega N +
\Omega b^{\dagger}b -\epsilon N^2\right)}D_m^{\dagger}(\eta N)
\end{equation}
We will consider the initial state of the field to be in a
coherent state
\begin{equation}
|\alpha\rangle   =
e^{-\frac{|\alpha|^2}{2}}\sum_{n=0}^{\infty}\frac{\alpha^n}{\sqrt{n!}}|n\rangle.
\end{equation}
and the initial state of the mirror to be arbitrary and denoted by
the density matrix $\rho_m$. We may calculate then $\langle
a\rangle$ in the form
\begin{equation}
\langle a\rangle=\alpha e^{-i(\omega+\epsilon)t} Tr \left[\rho_m
D_m\left( \eta e^{i\Omega t}\right)D_m\left(- \eta\right)|\alpha
e^{2i(\epsilon t- \eta^2\sin\Omega t)}\rangle
\langle\alpha|\right]
\end{equation}
where we have used several times the properties of permutation
under the trace symbol. By using that
\begin{equation}
D_m\left( \eta e^{i\Omega t}\right)D_m\left(- \eta
\right)e^{i\eta^2\sin\Omega t}= D_m\left( \eta(e^{i\Omega
t}-1)\right)
\end{equation}
we may finally write
\begin{equation}
\langle a\rangle=\alpha e^{-i(\omega+\epsilon)t}
e^{-i\eta^2\sin\Omega t} e^{-|\alpha|^2(\epsilon t-
\eta^2\sin\Omega t)} \chi_m\left( \eta(e^{i\Omega t}-1)\right)
\end{equation}
where $\chi_m\left( \eta(e^{i\Omega t}-1)\right)$ is the
characteristic function associated to the density matrix $\rho_m$.
Therefore, by measuring the quadratures  of the field (see for
instance \cite{leo}) $\langle
X\rangle=\langle(a+a^{\dagger})\rangle/\sqrt{2}$ and $\langle Y
\rangle =-i\langle(a-a^{\dagger})\rangle/\sqrt{2}$ we may obtain
the average value for the annihilation operator and hence,
information about the state of the mirror through its
characteristic function. The argument of the characteristic
function may be changed in some range of parameters as $\omega\sim
10^{16} s^{1}$,  $\Omega\sim 1$ kHz, $L\sim 1$ m and $m\sim 10 $
mg \cite{manci,mey1,mey2}. One could use the present method to
reconstruct the quantum superpositions of a mirror state recently
proposed by Marshall {\it et al.} around the origin to look for a
negative Wigner function in this region.

What makes it possible to obtain information about the mirror
state is the initial coherence of the field and the form of the
Hamiltonian that has the term $b+b^{\dagger}$. Wilkens and Meystre
\cite{wil} had shown that for the Jaynes-Cummings Model (JCM) (see
for instance \cite{jcm}) it was possible to obtain information
about the characteristic function of the field only if the system
interacted with an extra (classical) field to allow several
absorptions ($a^k$) or emissions [$(a^{\dagger})^k$]. The JCM by
itself would allow one emission or absorption at a time because of
the form of the interaction Hamiltonian
$H_I=\lambda(a\sigma_++\sigma_-a^{\dagger})$ where the $\sigma$'s
the usual spin operators and $\lambda$ the interaction constant.

However, if we do not make the rotating wave approximation in the
atom field interaction it was shown that transforming the complete
Hamiltonian by means of a unitary transformation gives
\cite{moyac}
\begin{equation}
H_T=\omega N + \omega_0
W\left(\frac{\lambda}{\omega}\sigma_z\right)
\end{equation}
where $W\left(\frac{\lambda}{\omega}\sigma_z\right) =
D\left(\frac{\lambda}{\omega}\sigma_z\right) (-1)^N D^{\dagger}
\left(\frac{\lambda}{\omega}\sigma_z\right)$ is the Wigner
operator \cite{vog}. This hints that keeping terms in the
Hamiltonian proportional to the sum of annihilation and creation
operators allows information about the system to be obtained.

In conclusion, we have shown that by measuring filed quadratures
one may be able to reconstruct the characteristic function for the
density matrix of the mirror.

We would like to thank CONACYT for support.

\end{document}